\documentclass[aps,pre,showkeys,showpacs,
reprint,longbibliography,
nofootinbib,
floatfix]{revtex4-1}
\usepackage[utf8]{inputenc}
\usepackage[T1]{fontenc}
\usepackage{amsmath}
\usepackage{amsfonts}
\usepackage{graphicx}
\usepackage{subcaption}
\usepackage{psfrag}
\usepackage[justification=centerlast]{caption}
\usepackage[colorlinks=true,hyperfootnotes=true,breaklinks=true]{hyperref}

\begin{document}
\title{Multi-choice opinion dynamics model based on Latan\'e theory}

\author{Przemysław Ba\'ncerowski}
\author{Krzysztof Malarz}
\homepage{http://home.agh.edu.pl/malarz/}
\email{malarz@agh.edu.pl}

\affiliation{\href{http://www.agh.edu.pl/}{AGH University of Science and Technology},
\href{http://www.pacs.agh.edu.pl/}{Faculty of Physics and Applied Computer Science},
al. Mickiewicza 30, 30-059 Krakow, Poland}

\date{\today}
\begin{abstract}
	In this paper Nowak--Szamrej-Latan\'e model is reconsidered. This computerised model of opinion formation bases on Latan\'e theory of social impact. We modify this model to allow for multi (more than two) opinions. With computer simulations we show that in the modified model the signatures of order/disorder phase transition are still observed. The transition may be observed in the average fraction of actors sharing the $i$-th opinion, its variation and also average number of clusters of actors with the same opinion and the average size of the largest cluster of actors sharing the same opinion. Also an influence of model control parameters on simulation results is shortly reviewed. For a homogeneous society with identical actors' supportiveness and persuasiveness the critical social temperature $T_C$ decreases with an increase of available opinions $K$ from $T_C=6.1$ ($K=2$) via 4.7, 4.1 to $T_C=3.6$ for $K=3$, 4, 5, respectively.
\end{abstract}

\date{\today}

\pacs{89.65.-s,	
89.75.-k}	

\keywords{Complex systems; Social and economic systems; Opinion dynamics; Ising and Potts model; Long range interactions}

\maketitle

\section{Introduction}

Simulations of opinion dynamics~\cite{Stauffer2009} are core subject of sociophysics~\cite{GalamReview,*GalamSociophysics}, an interdisciplinary field of research in complex systems directly connected to computational sociology.
Numerous examples of such research are published in interdisciplinary sections of physical journals~\cite{ISI:000415221300002,Nyczka2013,Kulakowski2009469,Gekle-2005,Sznajd-2005a,Amblard-2004,Holyst-2000,Kacperski-2000}, and in journals devoted to computational sociology \cite{Mathias-2016,Gronek2011,Deffuant-2006,Hegselmann-2002,ISI:000402671600001,ISI:000402671600005,ISI:000397168100012}.
The models of opinions dynamics deals with binary (or Boolean), Ising-like~\cite{Lenz1920,*Ising1925} variables, corresponding to two-states models of opinions~\cite{Kulakowski2008,Slanina-2008,Sznajd-2005,Sznajd-2000} or multi-state, but still discrete state opinions models \cite{Kulakowski2010,Gekle-2005} or discrete vector-like variables~\cite{Sznajd-2005a}.
The second group of models deals with continuous opinions~\cite{Mathias-2016,Deffuant-2006,Hegselmann-2002,Deffuant-2000,ISI:000411951900001,Malarz2006b,ISI:000413837700014,ISI:000414818100049,ISI:000409101900001,ISI:000399951700004,ISI:000399951000005,ISI:000396054300030}.

Another classification of opinion dynamics models may be based on geometry of underlying network of connections among actors.
Basing on this criteria we can deal with continuous (plane-like) \cite{Kulakowski2009469,Kulakowski2014,Kulakowski2012,Gronek2011} or discrete geometry.
The later may be divided into additional sub-groups, with regular lattices~\cite{Kulakowski2010,Sznajd-2005,Sznajd-2000,ISI:000415596700004,ISI:000411951900001} or complex networks~\cite{Kulakowski2008,Amblard-2004,ISI:000412648300002,ISI:000404700100004,ISI:000401622900002,ISI:000400473800014,ISI:000390102700002,ISI:000386744400041}.

The last classification includes system dynamics in terms of time evolution of the system, which again may occurring in discrete or in continuous time.
 
Assumed scheme of system representation force choosing the most adequate numerical technique for computer simulation of the system, including solving set of differential equations
 \cite{evans2010partial}
(continuous space of opinions, continuous geometry and continuous time) or cellular automata technique~\cite{Hegselmann-2000,*Ilachinski-2001,*Wolfram-2002,*Chopard-2005,*Chopard-2012} (discrete space of opinions, discrete time and discrete geometry).

In this paper Nowak--Szamrej--Latan\'e model is reconsidered~\cite{Nowak-1990}.
We propose multi-choice opinion dynamics model based on Latan\'e~\cite{Latane-1976,Latane-1981a,Latane-1981} theory.
With computer simulation we show that in the system with
	the long-range interactions among actors 
	and more than two opinions
the order-disorder phase transition is also observed.

\subsection{Theory of social impact}

\begin{figure*}[!htbp]
	\centering
	\begin{tabular}{rlll}
	& $t=0$ & $t=1$ & $t=10$ \\
	$T=$&
	\includegraphics[height=0.258\textwidth]{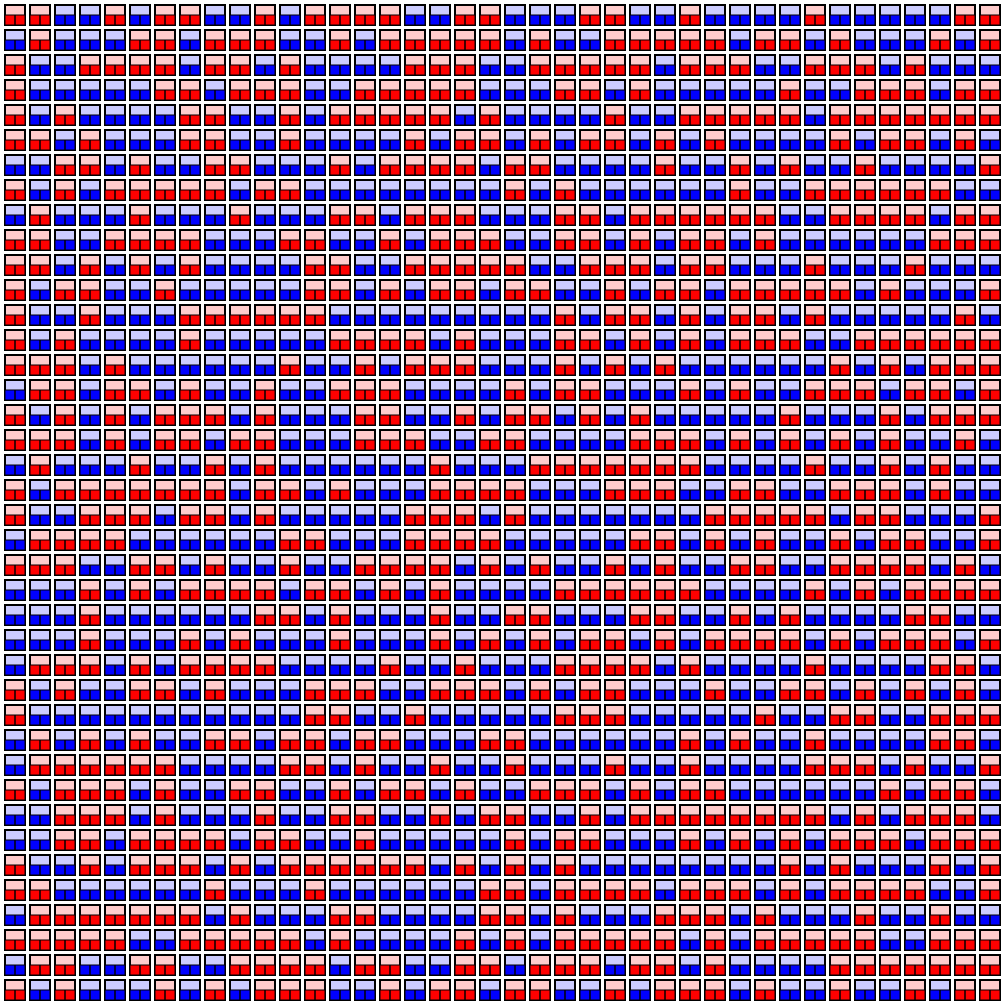} &
	\includegraphics[height=0.258\textwidth]{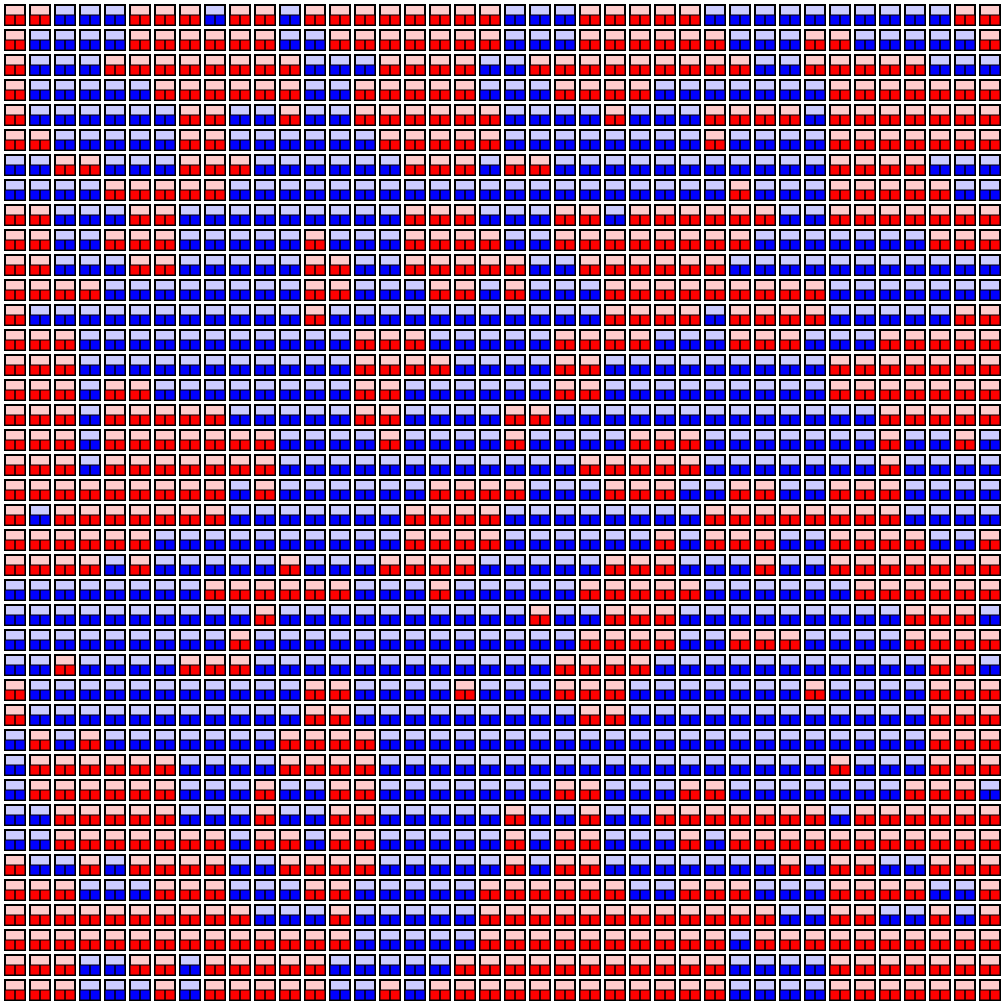} &
	\includegraphics[height=0.258\textwidth]{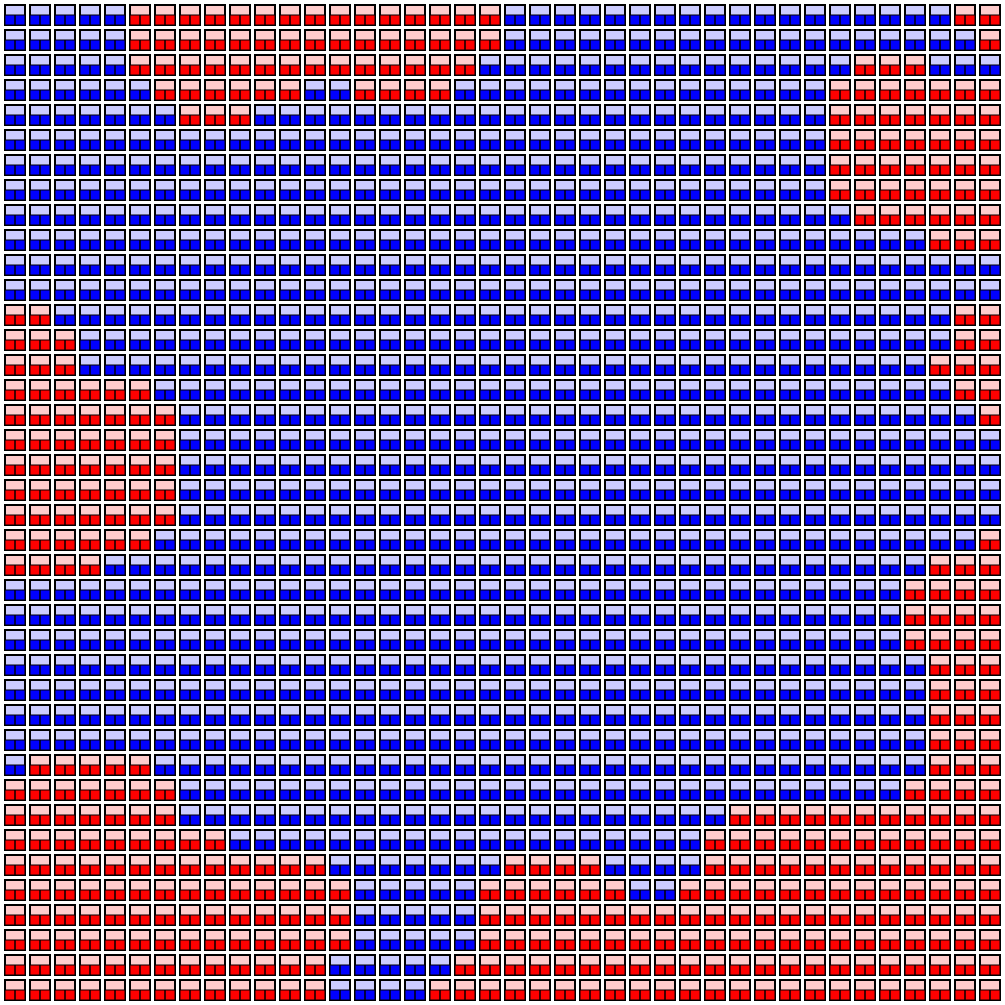}\\
	$0$ &
	\includegraphics[height=0.30\textwidth]{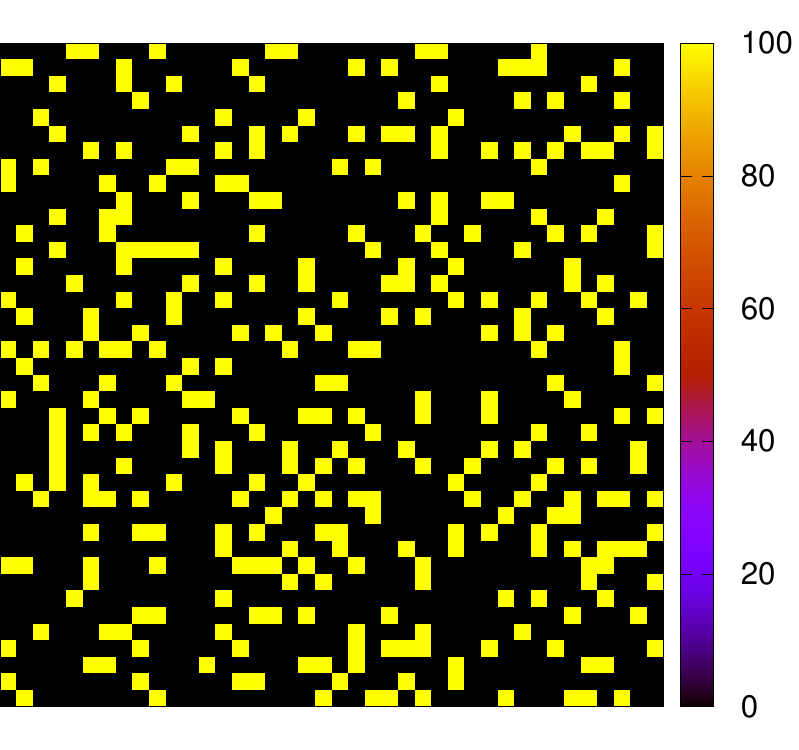}&
	\includegraphics[height=0.28\textwidth]{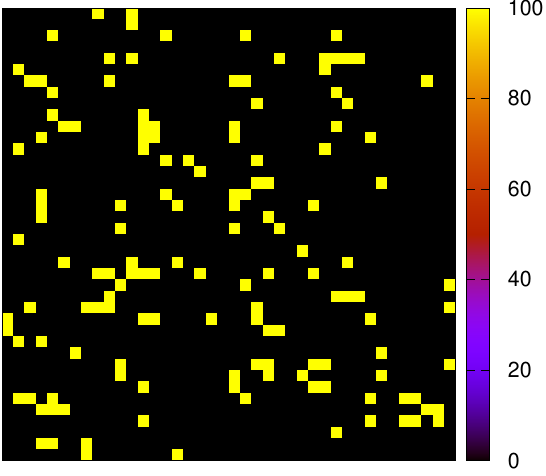}&
	\includegraphics[height=0.28\textwidth]{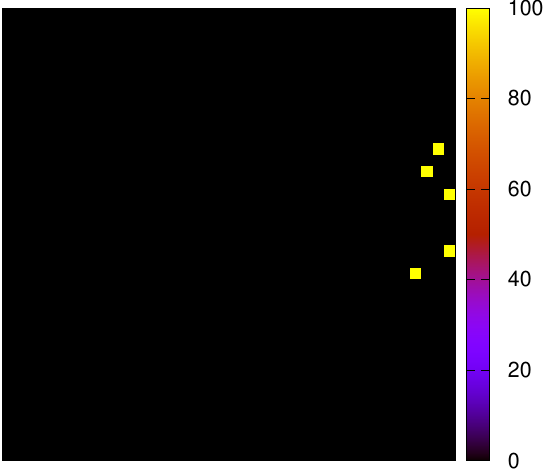}\\
	$1$ &
	\includegraphics[height=0.28\textwidth]{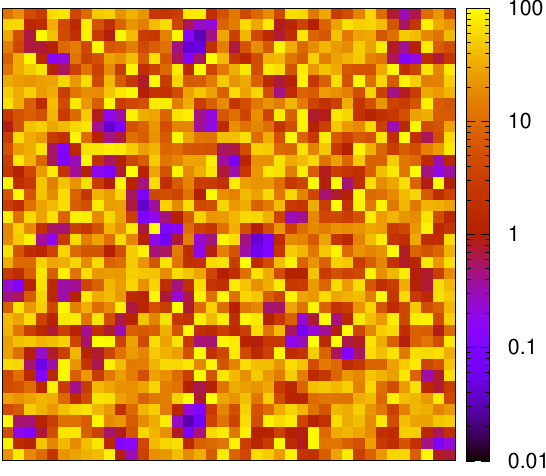}&
	\includegraphics[height=0.28\textwidth]{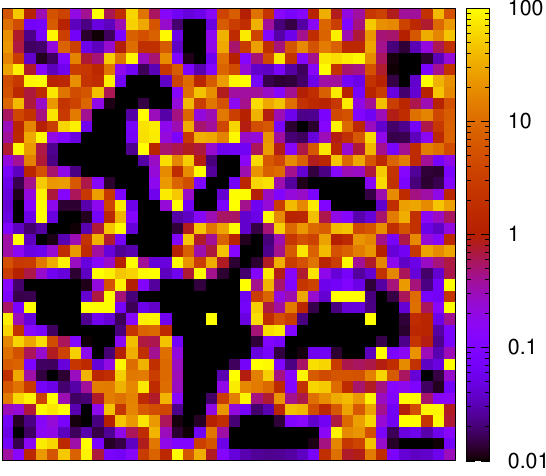}&
	\includegraphics[height=0.28\textwidth]{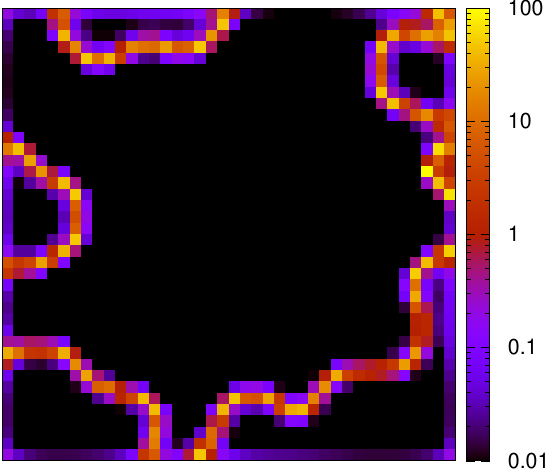}\\
	$3$ &
	\includegraphics[height=0.28\textwidth]{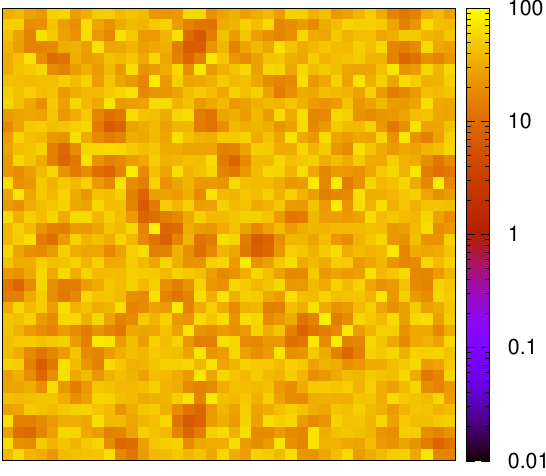}&
	\includegraphics[height=0.28\textwidth]{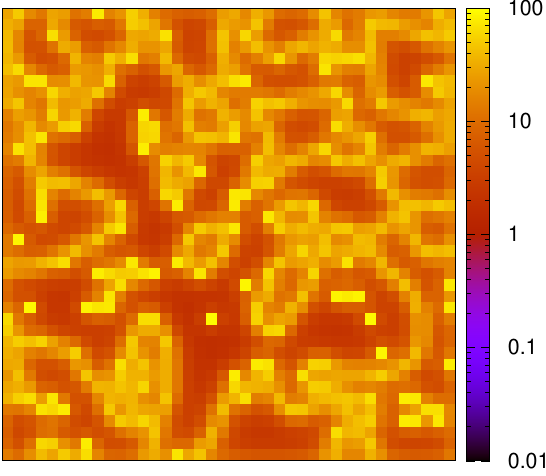}&
	\includegraphics[height=0.28\textwidth]{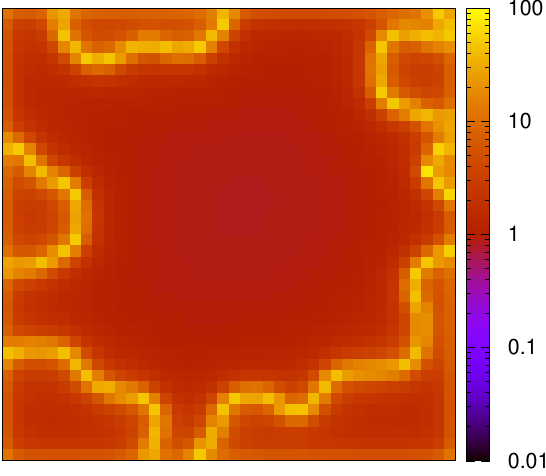}
	\end{tabular}
	\caption{\label{F:Pc}(Colour online) In the top line the snapshots from simulation of the system containing $L^2=40^2$ sites and $K=2$ are presented.
	The subsequent columns correspond to time steps $t=0$, 1, 10, respectively.
	In subsequent rows the probabilities of changing opinion $\mathcal{P}_i$ associated with sites $i$ and for social temperature $T=0$, 1 and 3 are presented.
	$\forall i: p_i=s_i=0.5$, $\alpha=3$~\cite{ThesisBancerowski}.}
\end{figure*}

The mathematical model being the foundation of this work relies on Latan\'e social impact theory~\cite{Latane-1976,Latane-1981a,Latane-1981} and its computerised version proposed by~\citet{Nowak-1990}.
This approach for binary opinions and possible charismatic leader localised in the system centre has been thoroughly explored in Ho{\l}yst, Kacperski and Schweitzer papers~\cite{Holyst-2000,Kacperski-2000} (see Ref.~\cite{ARCPIX253} for review).

Latan\'e assumes that people are social animals and in their natural environment (society) they influence each other.
These interactions do not have to be intentional.
Under this assumption we understand all interactions among people.
Persuasion, joke, sharing emotions and feelings---all of these can affect others.
Latan\'e describes these interactions as {\em social impact}.

The theory of social impact bases on three fundamental principles: {\it i}) social force, {\it ii}) psycho-social law and {\it iii}) multiplication/division of impact.

\subsubsection{Social force}

The social force principle \cite{Latane-1981} says that social impact $I$ on $i$-th actors is a function of the product of strength $S$, immediacy $J$, and the number of sources $N$
\begin{equation} I=\mathcal{F}(SJN). \label{eq:socialforce}\end{equation}

The strength of influence is the intensity, power or importance of the source of influence. This concept may reflect socio-economical status of the one that affects on our opinion, his/her age, prestige or position in the society.

The immediacy determines the relationship between the source and the goal of influence. This may mean closeness in the social relationship, lack of communication barriers and ease of communication among actors.

Latan\'e called this principle `a bulb theory of social relations'. 
According to this analogy the social impact plays a role of illuminance.
The illuminance depends on 
\begin{itemize}
\item the power of the bulb (physicists prefer to think about bulb's luminous flux)---equivalent of the strength of impact \item the distance from sources (bulbs)---equivalent of the immediacy \item and the number of bulbs---equivalent of the number of people.  
\end{itemize}

\subsubsection{Psycho-social law}

The data of the famous \citet{Asch1955} and \citet{Milgram1969} experiments may be fitted to formula proposed by Latan\'e: 
\begin{equation} I\propto SN^\beta, \label{eq:psychosocial}\end{equation}
where $N$ is the number of people exerting the impact, $S$ is a strength of impact and $0<\beta<1$ is the scaling exponent.
This means that each next actor $j$ sharing the same opinion as actor $i$ exerts the lower impact on the $i$-th actor.
This formula has been independently confirmed experimentally by \citet{Latane-1976}.

\subsubsection{Multiplication/division of impact}

The lecture for single student influence his/her much more the same lecture given for hundred of students.
In the latter case, the impact of lecture is roughly equally divided among all listeners~\cite{Darley1968}.
For this issue Latan\'e proposes
\begin{equation} I\propto SN^{-\gamma}, \label{eq:multidivision}\end{equation}
where the scaling exponent $0<\gamma<1$.
\citet{Latane-1981a} gathered results over hundred experiments to validate Eq.~\eqref{eq:multidivision}.

\begin{figure}[!htbp]
	\psfrag{t}{$T=$}
	\psfrag{i}{$t$}
	\psfrag{pc}{$\mathcal{\bar P}$ [\%]}
\includegraphics[width=.48\textwidth]{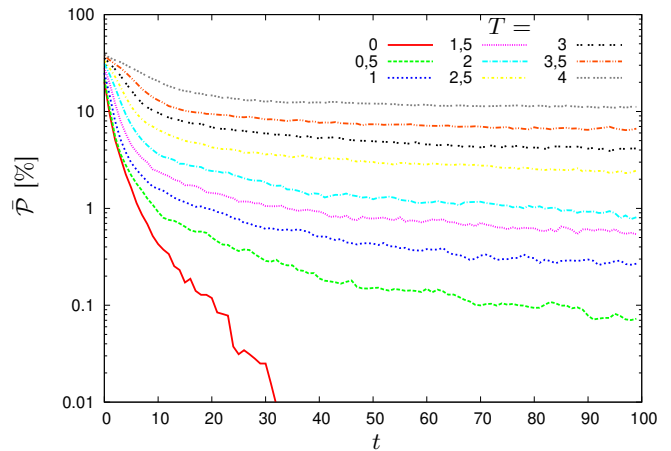}
	\caption{\label{F:avePc_vs_T} The time evolution of the average changing opinion probability $\mathcal{\bar P}$ [\%] for various values of social temperatures $T$ \cite{ThesisBancerowski}.}
\end{figure}

\subsubsection{The limitations of the theory}

The main limitation of the social impact theory lies in treating people as totally passive.
The second trouble is the absence of dynamics in the model.
These issues have been solved by~\citet{Nowak-1990} in the computerised version of Latan\'e model.

\section{\label{S:model}Model}

\begin{figure*}[!htbp]
\begin{tabular}{lll}
	& $t=0$ & $t=10$ \\
	& \includegraphics[height=0.295\textwidth]{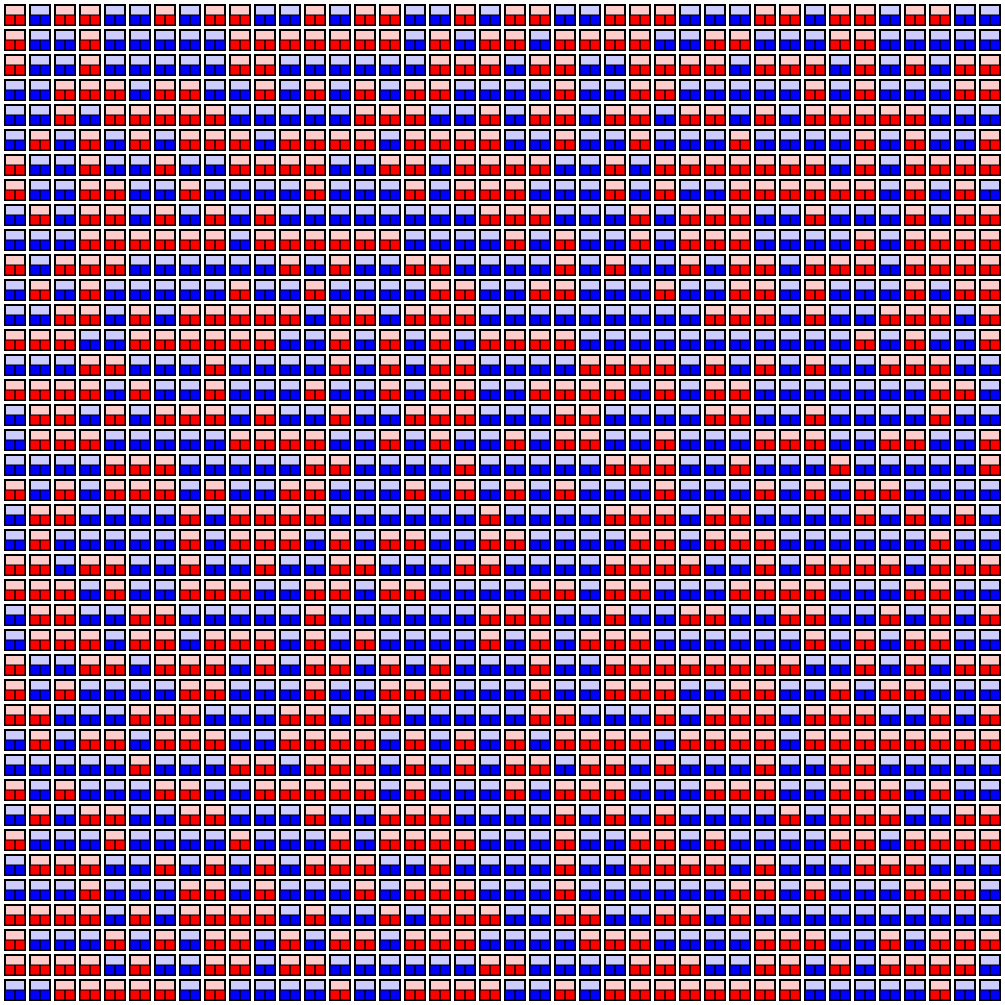} & \includegraphics[height=0.295\textwidth]{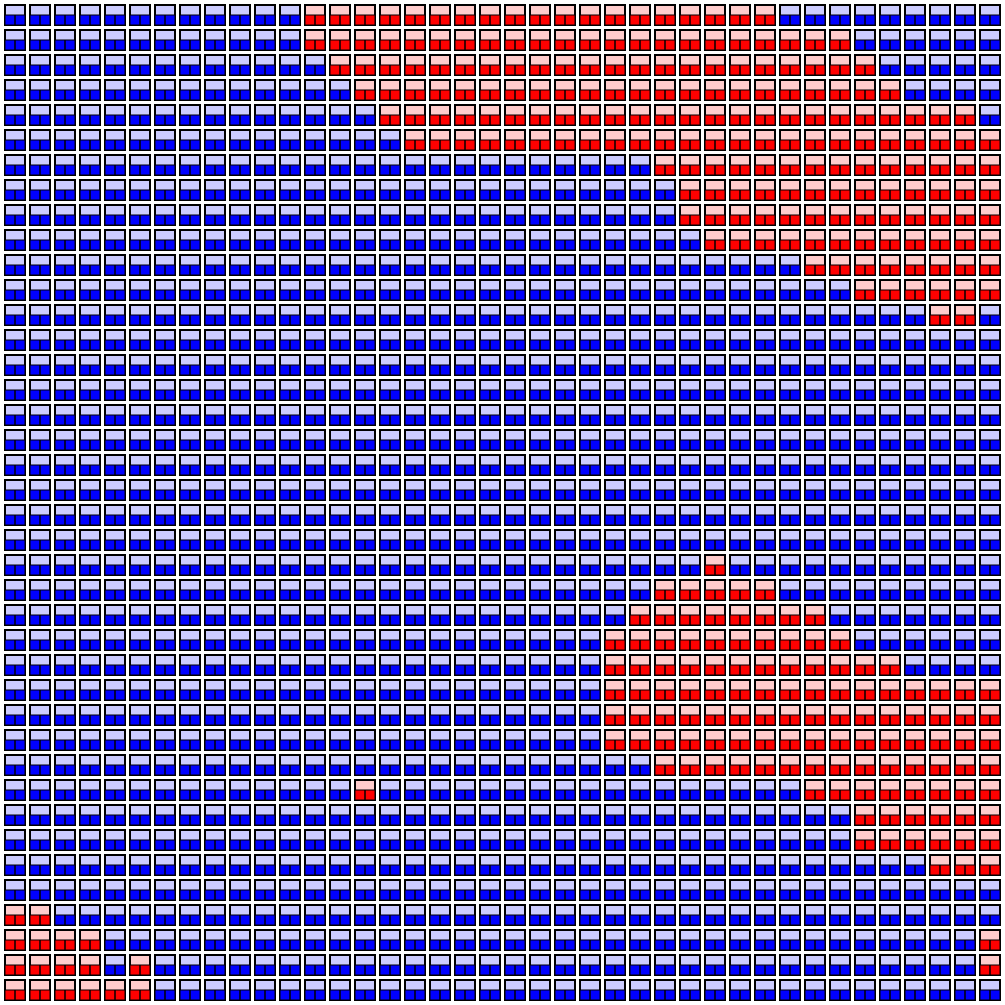}\\
	$\alpha=2$ & \includegraphics[height=0.31\textwidth]{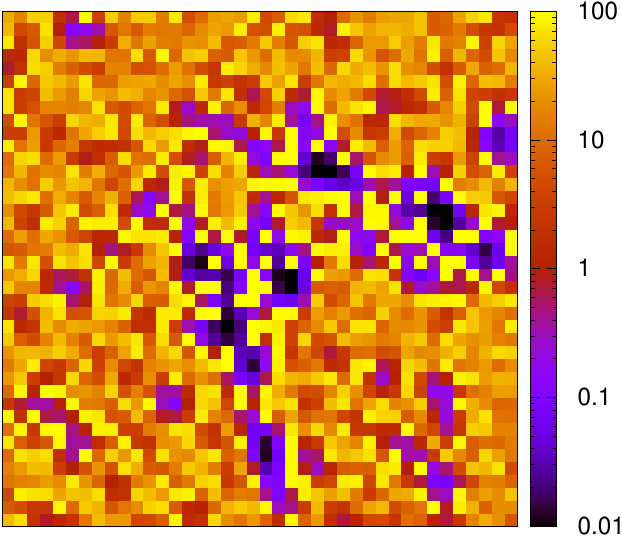} & \includegraphics[height=0.31\textwidth]{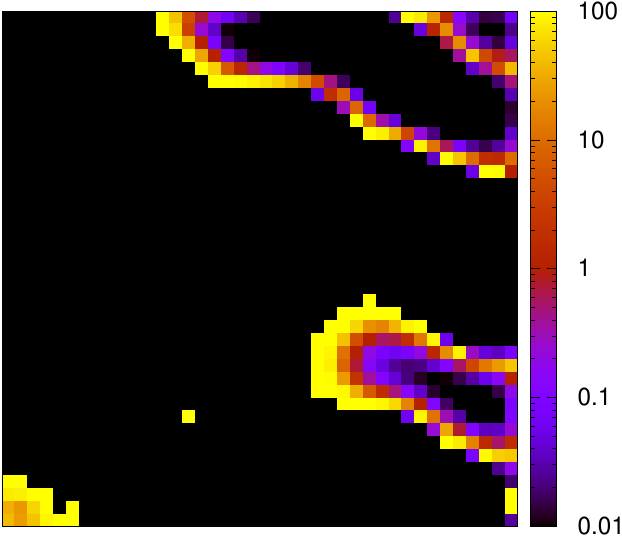}\\
	$\alpha=3$ & \includegraphics[height=0.31\textwidth]{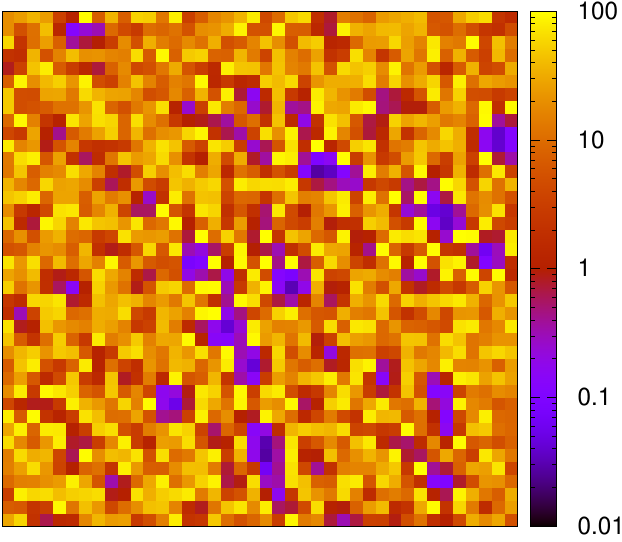} & \includegraphics[height=0.31\textwidth]{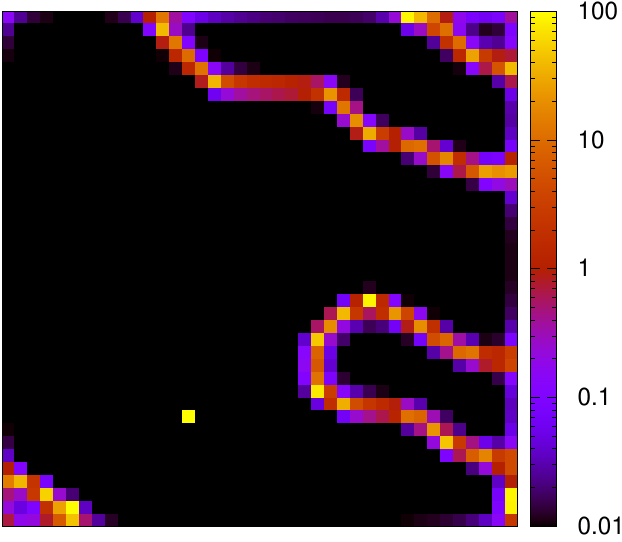}\\
	$\alpha=6$ & \includegraphics[height=0.31\textwidth]{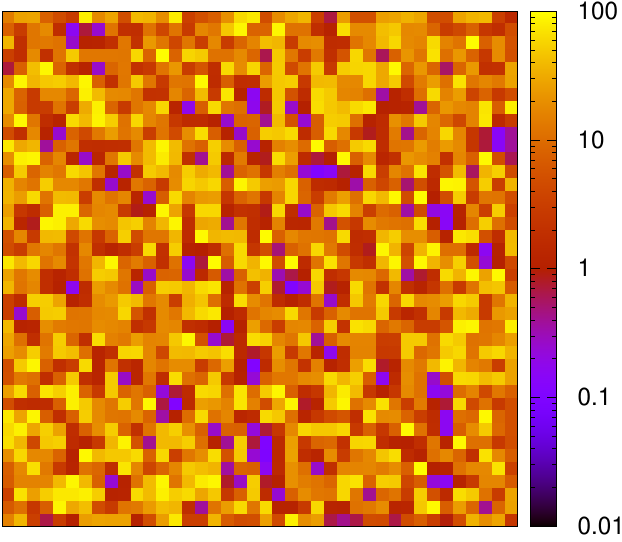} & \includegraphics[height=0.31\textwidth]{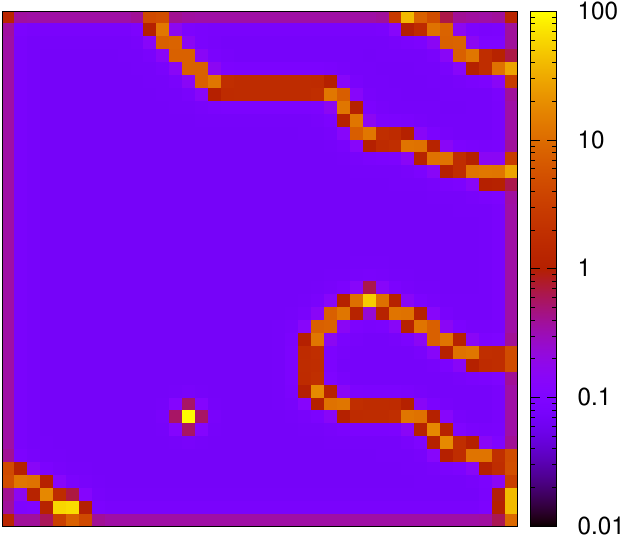}\\
\end{tabular}
	\caption{\label{F:Pc_vs_alpha}(Colour online) The maps of probabilities of opinion changes $\mathcal{P}_i$ [\%] for social temperature $T=1$ at the initial random distribution of opinions ($t=0$) and after ten time steps of simulation ($t=10$) and for various values of exponent $\alpha=2$, 3 and 6.
In the first row the snapshots from simulations indicating the spinsons opinions for $t=0$ (first column) and $t=10$ (second column) are presented \cite{ThesisBancerowski}.}
\end{figure*}

Every actor at position $i$ is characterised by his/her discrete {\em opinion} $\xi_i$, his/her {\em persuasiveness} ($0\le p_i\le 1$) and his/her {\em supportiveness} ($0\le s_i\le 1$).
Parameter $p_i$ describes the intensity of persuasion to change the opinion by actor $i$ from a person with opinion different than $\xi_i$, while $s_i$ describes the intensity of supporting people with the same views.

\subsection{\label{SM:two}Two opinions ($K=2$)}

For two opinions one can assume integer values of  $\xi_i\in\{-1,+1\}$.
For evaluation of social impact $I_i$ on actor at position $i$ one can apply formula proposed in Ref.~\cite{ARCPIX253}:
\begin{subequations}
\label{eq:holyst}
\begin{eqnarray}
	I_i(t) & = \mathcal{J}_P\left(\sum_{j=1}^{N}\dfrac{q(p_j)}{g(d_{i,j})}[1-\xi_i(t)\xi_j(t)]\right)\\
	       & - \mathcal{J}_S\left(\sum_{j=1}^{N}\dfrac{q(s_j)}{g(d_{i,j})}[1+\xi_i(t)\xi_j(t)]\right),
\end{eqnarray}
\end{subequations}
where $\mathcal{J}_P(\cdot)$, $\mathcal{J}_S(\cdot)$, $q(\cdot)$, $g(\cdot)$ stand for scaling functions and $d_{i,j}$ is Euclidean distance between sites $i$ and $j$.
The system dynamics may be governed by heat-bath-like dynamics~\cite{Kacperski-2000}, i.e.:
\begin{equation}
\label{eq-rule-2}
\xi_i(t+1)=
\begin{cases}
	 \xi_i(t) &\text{with probability } \dfrac{\exp\left(\dfrac{-I_i(t)}{T}\right)}{2\cosh\left(\dfrac{I_i(t)}{T}\right)},\\
	          & \\
	-\xi_i(t) &\text{with probability } \dfrac{\exp\left(\dfrac{ I_i(t)}{T}\right)}{2\cosh\left(\dfrac{I_i(t)}{T}\right)},
\end{cases}
\end{equation}
where $T$ is a noise parameter (social temperature~\cite{social_temperature}).

For $T=0$ the rule~\eqref{eq-rule-2} may be reduced to fully deterministic rule~\cite{Kacperski-2000}
\begin{equation}
	\label{eq:detemin}
	\xi_i(t+1)=\text{sgn}(I_i(t))
\end{equation}
as $I_i(t)=0$ is practically impossible to occur.

\subsection{\label{SM:multi}Three and more opinions ($K>2$)}

For multi-state space of opinions we do not assign numeric values to opinions $\xi_i\in\{\Xi_1,\Xi_2,\cdots,\Xi_K\}$, where $K$ is the number of available opinions.
We rather prefer to think about various `colours' of opinions, or about $K$ orthogonal versors in $K$-dimensional vector space.
Also we propose some modifications of Eq.~\eqref{eq:holyst}. 
We propose to separate the social impact on actor $i$ from actors $j$ sharing opinion of actor $i$ ($\xi_j=\xi_i$)
\begin{subequations}
\label{eq:wplyw}
\begin{equation}
	I_{i,k}(t) = 4\mathcal{J}_s\left(\sum_{j=1\atop \xi_j=\xi_i}^{N}\dfrac{q(s_j)}{g(d_{i,j})}\right)
\label{eq:wplyw_ta_sama}
\end{equation}
and all other actors having different $K-1$ opinions ($\xi_j\ne\xi_i$)
\begin{equation}
	I_{i,k}(t) = 4\mathcal{J}_p\left(\sum_{j=1 \atop \xi_j=k\neq\xi_i}^{N}\dfrac{q(p_j)}{g(d_{i,j})}\right),
\label{eq:wplyw_inna}
\end{equation}
\end{subequations}
where $1\le k\le K$ enumerates the opinions.
The factor of four in Eq.~\eqref{eq:wplyw} guaranties exactly the same impact on actor $i$ as calculated basing on Eq.~\eqref{eq:holyst} for $K=2$.

The calculated social impacts $I_{i,k}(t)$ influence the $i$-th actor opinion $\xi_i(t+1)$ at the subsequent time step.
For $T=0$ this opinion is determined by those opinions which believers exert the largest social impact on $i$-th actor
\begin{multline}
	\xi_i(t+1)=\Xi_k \iff \\ I_{i,k}(t)=\max(I_{i,1}(t), I_{i,2}(t),\cdots,I_{i,K}(t)).
\end{multline}

For finite values of social temperature $T>0$ we apply the Boltzmann choice
\begin{equation}
	p_{i,k}(t) = \exp\left(\dfrac{I_{i,k}(t)}{T}\right),
\label{eq:E_ik}
\end{equation}
which yield probabilities
\begin{equation}
	P_{i,k}(t) = \frac{p_{i,k}(t)}{\sum_{j=1}^{K}p_{i,j}(t)}
\label{eq:P_ik}
\end{equation}
of choosing by $i$-th actor in the next time step $k$-th opinion:
\begin{equation}
	\xi_i(t+1)=\Xi_k, \text{ with probability } P_{i,k}(t).
\end{equation}
The form of dependence~\eqref{eq:E_ik} in statistics and economy is called logit function~\cite{Anderson1992,*Byrka-2016}.

We assume identity function for scaling functions $\mathcal{J}_S(x)\equiv x$, $\mathcal{J}_P(x)\equiv x$, $q(x)\equiv x$.
The distance scaling function should be an increasing function of its argument.
Here, we assume the distance scaling function as
\begin{equation}
\label{eq:g}
g(x)=1+x^\alpha,
\end{equation}
what ensures non-zero values $g(0)=1$ of denominator for self-supportivenees in Eq.~\eqref{eq:wplyw_ta_sama}.
Newly evaluated opinions are applied synchronously to all actors.

The simulations are carried out on square lattice of linear size $L=40$ with open boundary conditions.
We assume identical values of supportivenees and persuasiveness for all actors $\forall i: s_i=p_i=0.5$.
We set exponent $\alpha=3$ in the distance scaling function \eqref{eq:g}.

The web application allowing for direct observation of the system evolution is available at \url{http://www.zis.agh.edu.pl/app/MSc/Przemyslaw_Bancerowski/}.
The short manual for this application is available in Appendix~\ref{app:manual}.

\section{\label{S:results}Results}

\begin{figure}[!htbp]
	\psfrag{a}{$\alpha=$}
	\psfrag{i}{$t$}
	\psfrag{pc}{$\mathcal{\bar P}$ [\%]}
\centering
$T=0$\\
\includegraphics[width=.48\textwidth]{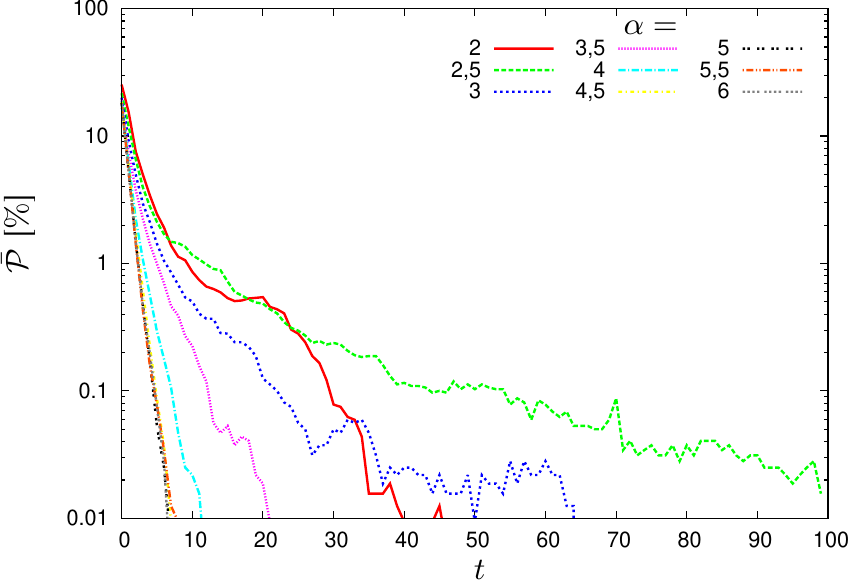}\\
$T=1$\\
\includegraphics[width=.48\textwidth]{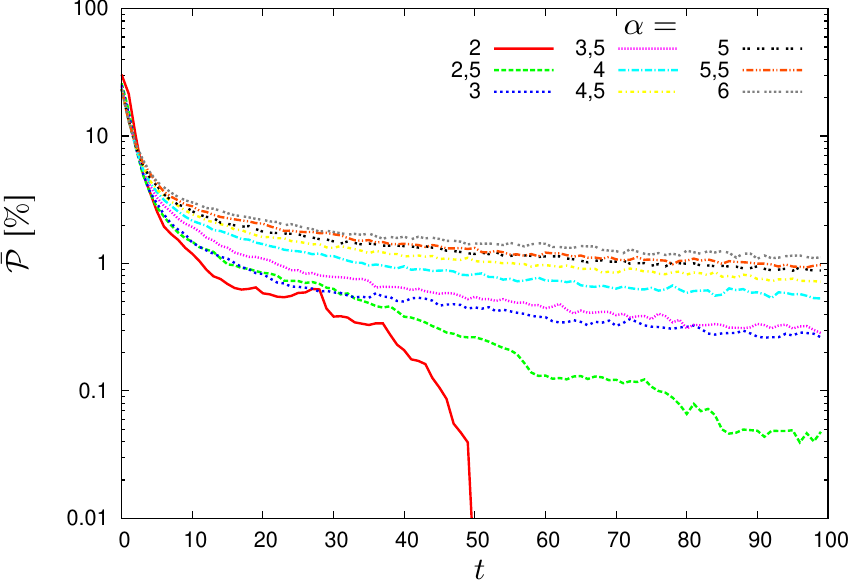}
	\caption{\label{F:avePc_vs_alpha} The time evolution of the average probability $\mathcal{\bar P}$ [\%] of opinion changes for various values of the distance function scaling exponents $\alpha$ \cite{ThesisBancerowski}.}
\end{figure}

\subsection{\label{SR:two}Two opinions}

\subsubsection{Influence of the model parameters on opinion dynamics}

To understand better the system time evolution the maps of probabilities $\mathcal{P}_i$ of opinion changes at sites $i$ (for $K=2$) are presented in Fig.~\ref{F:Pc}. 
The snapshots of system states at $t=0$, 1 and 10 are presented in the first row of Fig.~\ref{F:Pc}.
The corresponding to these states probabilities of opinion changing (flipping) for social temperatures $T=0$, 1 and 3 are presented in the second, third and fourth row of the Fig.~\ref{F:Pc}, respectively.
For $T=0$ (the second row) the system is fully deterministic and $\forall i: \mathcal{P}_i\in\{0,1\}$.
For long enough times of evolution the system reaches the nearly-steady state (with single spinsons\footnote{The term `spinson' comes from merging words `spin' and `person' and it describes actor who may have only two opinions. The term was introduced in 2013 by~\citet{Nyczka2013}.} going to change their minds) and clearly defined borders between groups (clusters) of spinsons with different opinions.
The static picture of the system is also observed for $T>0$, with non-zero probabilities of changing opinions $\mathcal{P}_i$ for spinson $i$ located at the clusters borders.

In Fig.~\ref{F:avePc_vs_T} the time evolution of the spatial average of probabilities of opinion changes
\[
	\mathcal{\bar P}=L^{-2}\sum_{i=1}^{L^2}\mathcal{P}_i
\]
is presented.
The spatial average over $L^2$ sites is marked through this paper by a bar ($\bar\cdot$).
For long enough times the average probabilities of opinion changes $\mathcal{\bar P}$ increases smoothly with increase of social temperature, reaching $\mathcal{\bar P}\approx 10\%$ for $T=4$.

As expected, an increase the social temperature $T$ enhances the spinsons nonconformity, i.e. they are able to change their minds although social impact exerted on them by other members of the society with the same opinion.
In the limit of infinite social temperature every actor chooses his/her opinion randomly, as
\[
	\lim_{T\to\infty} p_{i,k}(t)=1 \quad \text{ and } \quad \lim_{T\to\infty} P_{i,k}(t)=1/K.
\]

In Fig.~\ref{F:Pc_vs_alpha} the maps of probabilities changes $\mathcal{P}_i$ are presented again.
The first row shows the snapshots from simulations indicating the spinsons opinions for $t=0$ (first column) and $t=10$ (second column).
The subsequent rows correspond to probabilities of opinion changes for various values of exponent $\alpha$ in the distance scaling function $g(x)$ [Eq.~\eqref{eq:g}]---$\alpha=2$, 3, 6 in the second, third and fourth row, respectively.
The random initial configuration of opinions leads to random maps of $\mathcal{P}_i$. However, ten time steps of system relaxation allows for an observation of both: the spatial clusterization of spinsons shearing the same opinion and high probabilities of opinion changing at the borders of these clusters.
Moreover, for high values of exponent $\alpha$ differences among the minimal and the maximal values of $\mathcal{P}_i$ are much smaller than for small values of $\alpha$.

Quantitatively these differences may be observed in Fig.~\ref{F:avePc_vs_alpha} for purely deterministic ($T=0$) and non-deterministic ($T=1$) cases.
In principle, for $T>0$ the higher value of the exponent $\alpha$ leads to the higher value of $\mathcal{\bar P}$ which values saturate on the level $\mathcal{\bar P}\approx 1\%$ after hundred simulation steps for $T=1$ and $\alpha>4$.

\subsubsection{Phase transition}

\begin{figure}
\centering
\psfrag{i}{$t$}
\psfrag{t}{$T=$}
\psfrag{o}{$\bar\xi$}
\begin{subfigure}[b]{0.490\textwidth} \caption{\label{F:2op_s_sbar_t_a} $K=2$}
\includegraphics[width=.98\textwidth]{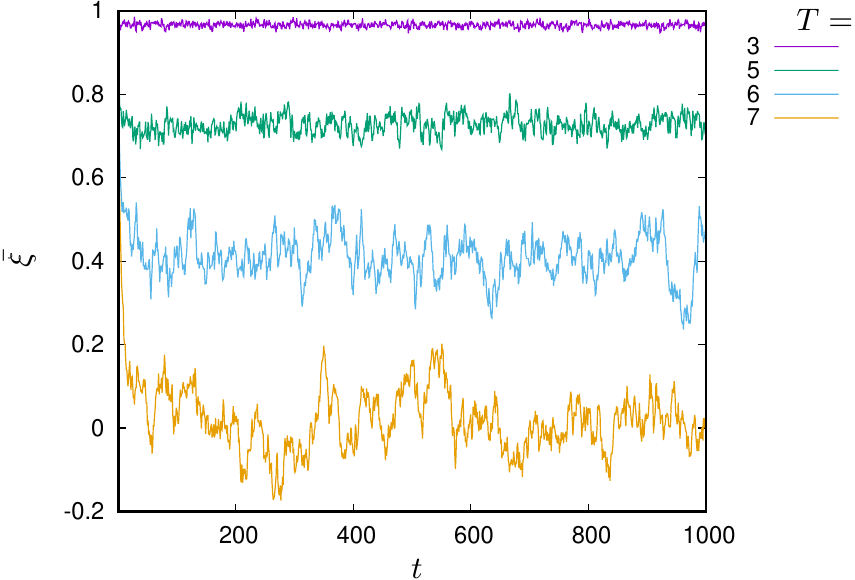}
\end{subfigure}
\begin{subfigure}[b]{0.490\textwidth} \caption{\label{F:2op_s_sbar_t_b} $K=2$, $T=6.2>T_C$}
\includegraphics[width=.98\textwidth]{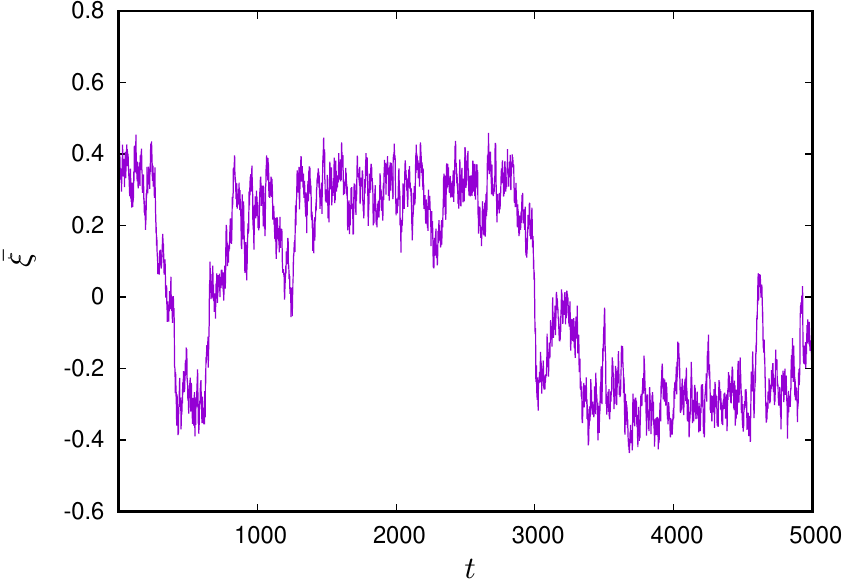}
\end{subfigure}
\caption{\label{F:2op_s_sbar_t}The time evolution of average opinion $\bar\xi$ for $K=2$ opinions and various social temperatures $T$.
$L=40$, $\alpha=3$, $\forall i:p_i=s_i=0.5$ \cite{ThesisBancerowski}.}
\end{figure}

In Fig.~\ref{F:2op_s_sbar_t_a} the results on an average opinion
\begin{equation} 
	\label{barxi}
	\bar\xi(t)\equiv L^{-2} \sum_{i=1}^{L^2}\xi_i(t), 
\end{equation}
for various values of the social temperature $T$ are presented.
Similarly to the Ising model some signatures of the phase transition in the system may be observed.
For low social temperature ($T<T_C$) the system is in ordered phase with majority of one (initially dominant) opinion.
However, for high enough temperature ($T>T_C$) the average opinion oscillates around $\bar\xi=0$.

In Fig.~\ref{F:2op_s_sbar_t_b} an example of time evolution of the spatial average opinion in the system for $T\to T_C^+$ is presented.
Although the long-range interaction among actors is assumed, the time evolution $\bar\xi(t)$ is not different from `magnetisation' evolution in the Ising model with characteristic `magnetisation' switching between its positive and negative values above the Curie temperature.

\begin{figure}
\centering
\psfrag{(a)}[][c]{(a) $K=2$}
\psfrag{(b)}[][c]{(b) $K=2$}
\psfrag{t}{$T$}
\psfrag{o}[][c]{$\langle\bar\xi\rangle$}
\psfrag{od}[][c]{$\sigma(\bar\xi)$}
\begin{subfigure}[b]{0.490\textwidth} \caption{\label{F:2op_save_sigma_T_a} $K=2$}
\includegraphics[width=.98\textwidth]{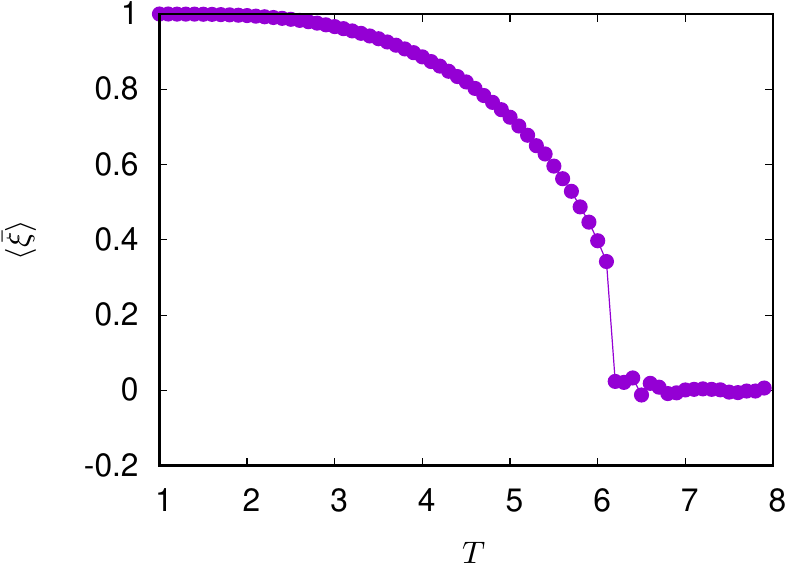}
\end{subfigure}
\begin{subfigure}[b]{0.490\textwidth} \caption{\label{F:2op_save_sigma_T_b} $K=2$}
\includegraphics[width=.98\textwidth]{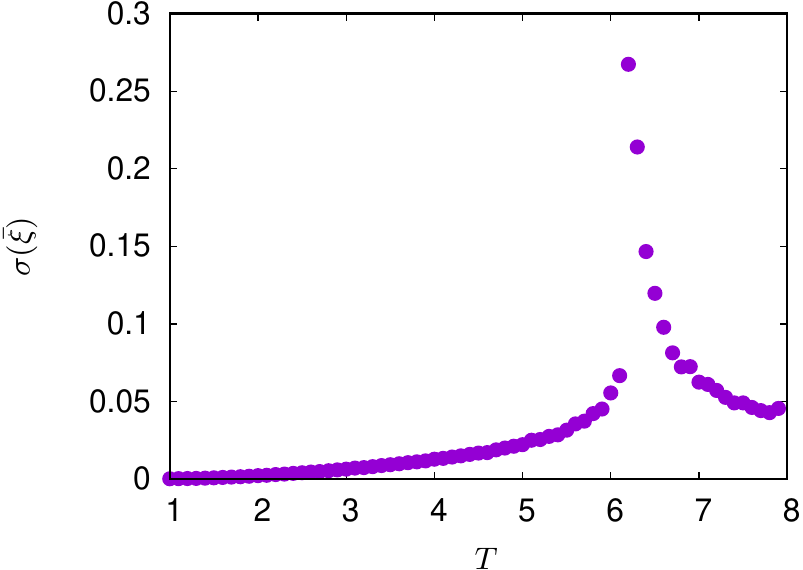}
\end{subfigure}
\caption{\label{F:2op_save_sigma_T} The values of (a) average values of opinion $\langle\bar\xi\rangle$ and (b) its standard deviation $\sigma(\bar\xi)$.
$L=40$, $\alpha=3$, $\forall i:p_i=s_i=0.5$.
The values of $\bar\xi$ are averaged over last $\tau=5000-100$ time steps \cite{ThesisBancerowski}.}
\end{figure}

In Fig.~\ref{F:2op_save_sigma_T_a} we plot the temporal average
\begin{equation}
	\label{avebarxi}
	\langle\bar\xi\rangle\equiv\tau^{-1} \sum_{t=t_0}^{t_M} \bar\xi(t),
\end{equation} for various temperatures $T$.
The temporal average over $\tau$ times steps is marked through this paper by brackets ($\langle\cdots\rangle$).
Here, $\tau=5000-100$, i.e. the first hundred of time steps is excluded from the averaging procedure.

The ordered phase phase vanishes for $T>T_C\approx 6.1$.
This critical value of $T_C$ coincidences nicely with a peak of average opinion dispersion 
\begin{equation}
	\label{sigma}
	\sigma^2(\bar\xi)=\langle{\bar\xi}\,^2\rangle-\langle\bar\xi\rangle^2
\end{equation}
as presented in Fig.~\ref{F:2op_save_sigma_T_b}.
The values of $\sigma$ plays a role of static susceptibility $\chi$ in Ising-like systems.
We confirm the earlier results indicating the phase transition in Nowak--Szamrej--Latan\'e model for binary opinions \cite{Holyst-2000}.

In the next Section we show that the above mentioned results are generic also when multi-opinions are available in the system. 

\subsection{\label{SR:multi}Three and more opinions}

\begin{figure*}[!htbp]
	\begin{subfigure}[b]{0.490\textwidth} \caption{\label{F:K3T0} $K=3$, $T=0$}
	\includegraphics[width=\textwidth]{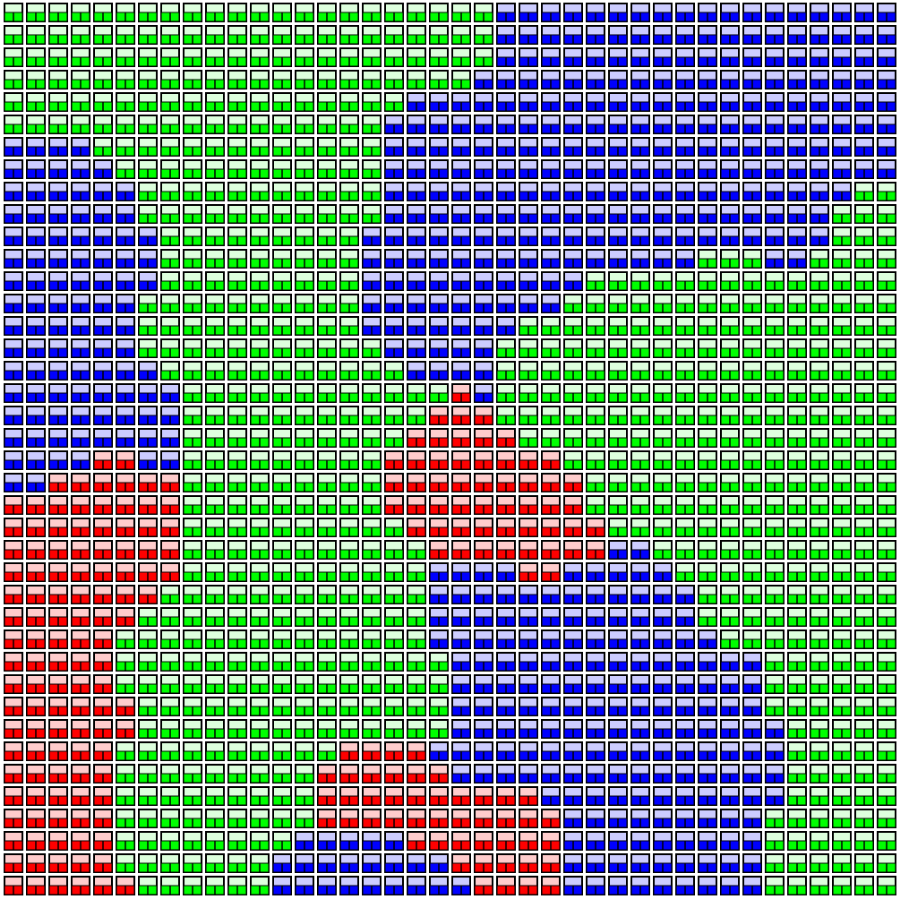} 
	\end{subfigure}
	\begin{subfigure}[b]{0.490\textwidth} \caption{\label{F:K3T6} $K=3$, $T=6$}
	\includegraphics[width=\textwidth]{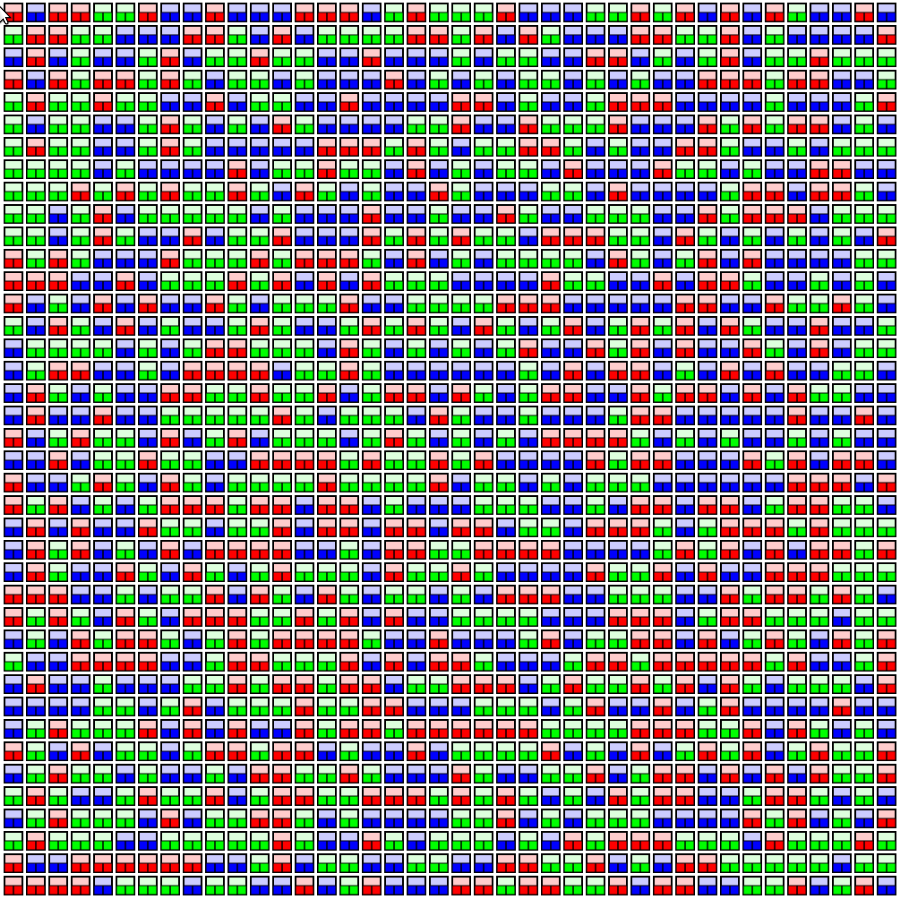} \\
	\end{subfigure}
	\begin{subfigure}[b]{0.490\textwidth} \caption{\label{F:K6T0} $K=6$, $T=0$}
	\includegraphics[width=\textwidth]{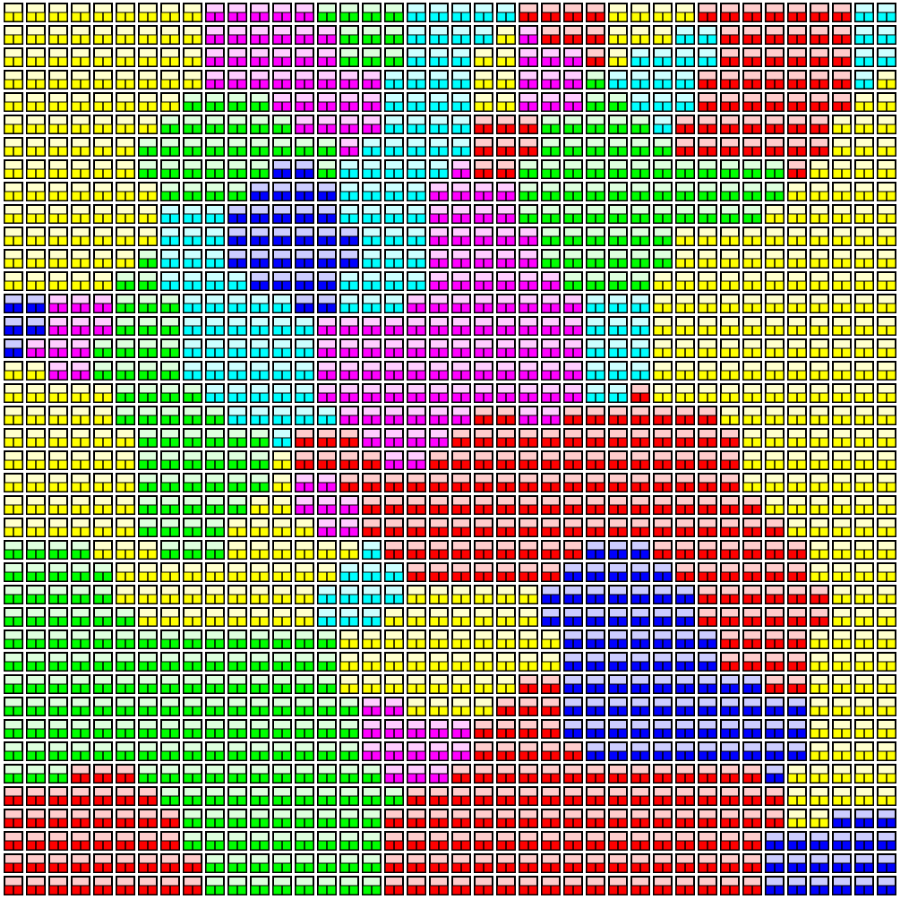} 
	\end{subfigure}
	\begin{subfigure}[b]{0.490\textwidth} \caption{\label{F:K6T6} $K=6$, $T=6$}
	\includegraphics[width=\textwidth]{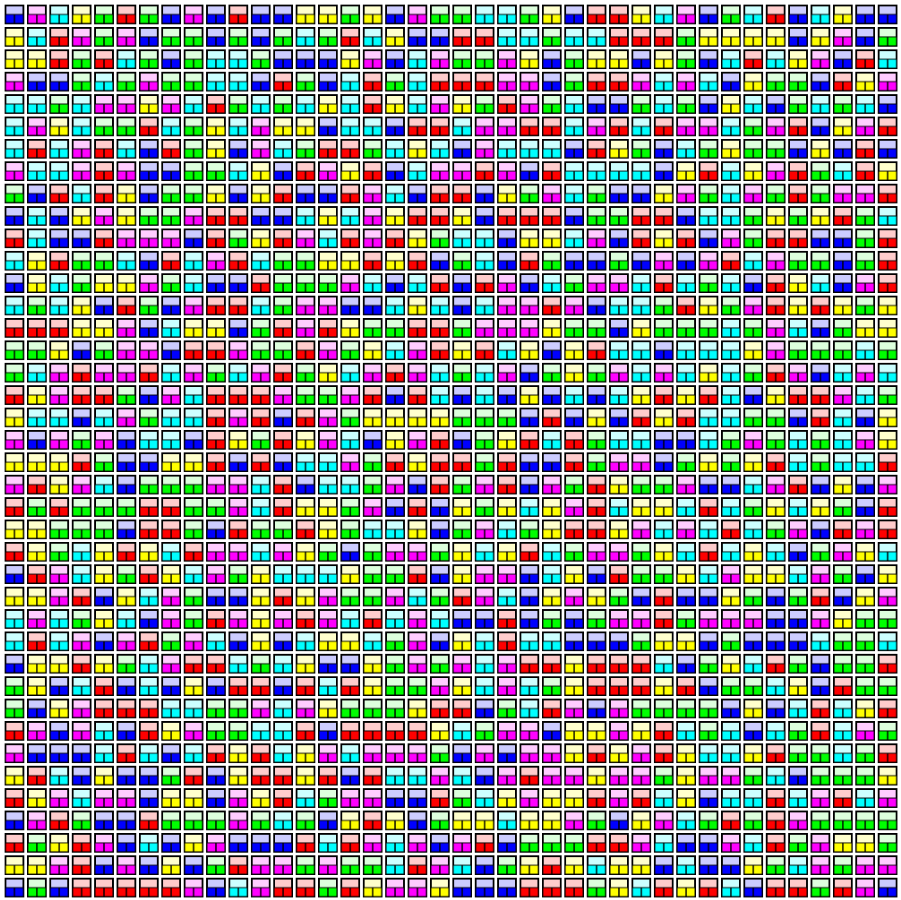}\\
	\end{subfigure}
\caption{\label{F:K3K6}(Colour online) The snapshots of opinions spatial distribution for social temperature $T=0$ (the first column) and $T=6>T_C$ (the second column) for various numbers of available opinions $K=3$ (the first row) and $K=6$ (the second row).}
\end{figure*}

As we mentioned in Section~\ref{SM:multi} for $K>2$ we do not assign numerical values to opinions $\xi_i$.
Instead, we prefer to think about $K$ `colours' $\Xi_{k=1,\cdots,K}$ of opinions (see Fig.~\ref{F:K3K6} for snapshots from simulations presenting spatial distributions of opinions for $T=0$, 6 and $K=3$, 6).
This assumption does not allow for dealing with $\langle\bar\xi\rangle$ [Eq.~\eqref{avebarxi}] and $\sigma(\bar\xi)$ [Eq.~\eqref{sigma}] in order to identify the critical social temperature $T_C$.
Thus for this purpose we propose to deal with a fraction $\langle\bar n_k\rangle$ of actors sharing the $k$-th opinion and its standard deviation $\sigma(\bar n_k)$. 

\begin{figure*}
	\centering
	\psfrag{A}{$\Xi_1$}
	\psfrag{B}{$\Xi_2$}
	\psfrag{C}{$\Xi_3$}
	\psfrag{D}{$\Xi_4$}
	\psfrag{E}{$\Xi_5$}
	\psfrag{avsize}[][c]{$\langle\mathcal{\bar S}_{\text{max}}\rangle$}
	\psfrag{s}[][c]{$\sigma(\bar n_k)$}
	\psfrag{nc}[][c]{$\langle\mathcal{\bar C}\rangle$}
	\psfrag{n}{$\langle\bar n_k\rangle$}
	\psfrag{T}{$T$}
	\begin{subfigure}[b]{0.329\textwidth} \caption{\label{F:345op_save_sigma_T_a} $K=3$}
	\includegraphics[width=\textwidth]{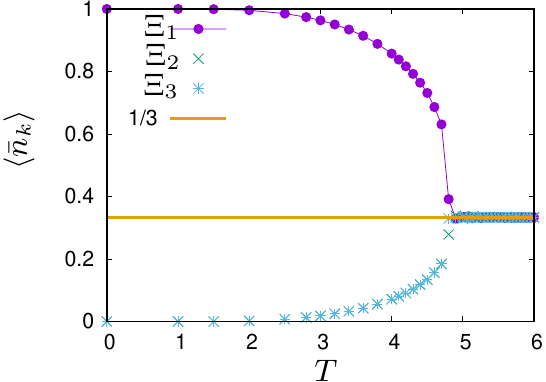}
	\end{subfigure}
	\begin{subfigure}[b]{0.329\textwidth} \caption{\label{F:345op_save_sigma_T_b} $K=4$}
	\includegraphics[width=\textwidth]{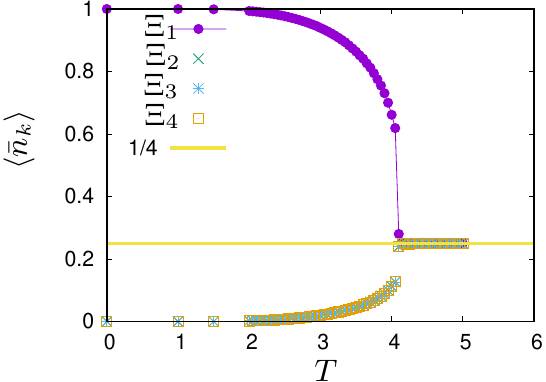}
	\end{subfigure}
	\begin{subfigure}[b]{0.329\textwidth} \caption{\label{F:345op_save_sigma_T_c} $K=5$}
	\includegraphics[width=\textwidth]{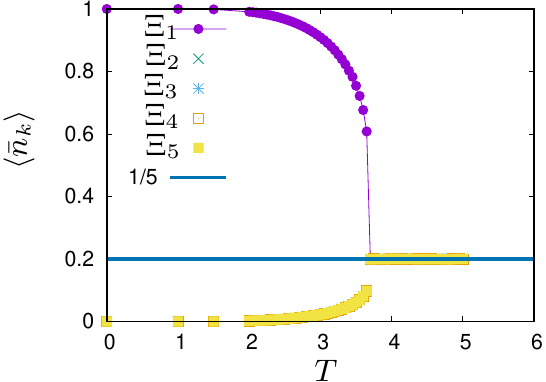}
	\end{subfigure}
	\begin{subfigure}[b]{0.329\textwidth} \caption{\label{F:345op_save_sigma_T_d} $K=3$}
	\includegraphics[width=\textwidth]{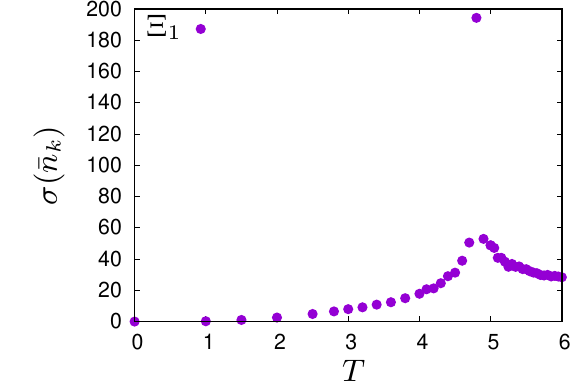}
	\end{subfigure}
	\begin{subfigure}[b]{0.329\textwidth} \caption{\label{F:345op_save_sigma_T_e} $K=4$}
	\includegraphics[width=\textwidth]{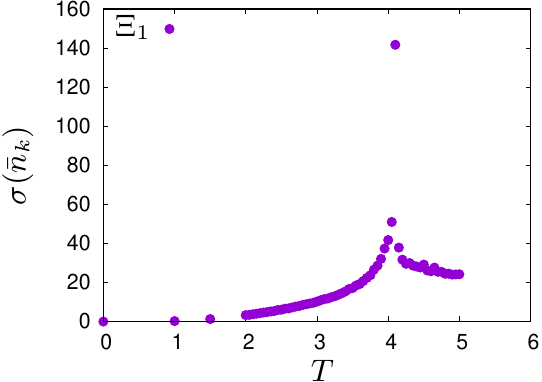}
	\end{subfigure}
	\begin{subfigure}[b]{0.329\textwidth} \caption{\label{F:345op_save_sigma_T_f} $K=5$}
	\includegraphics[width=\textwidth]{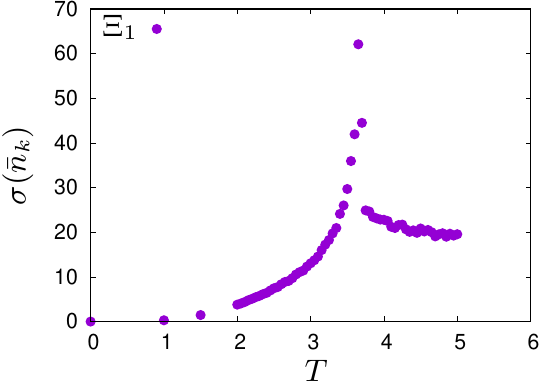}
	\end{subfigure}
	\begin{subfigure}[b]{0.329\textwidth} \caption{\label{F:345op_save_sigma_T_g} $K=3$}
	\includegraphics[width=\textwidth]{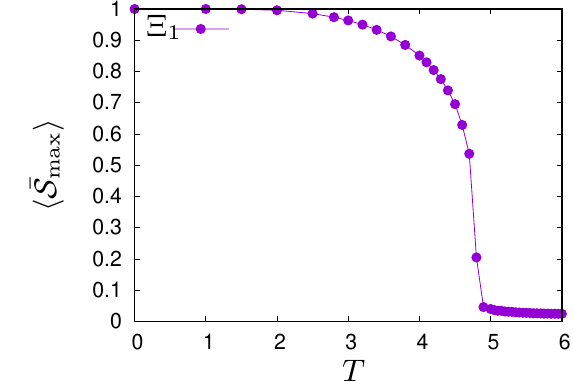}
	\end{subfigure}
	\begin{subfigure}[b]{0.329\textwidth} \caption{\label{F:345op_save_sigma_T_h} $K=4$}
	\includegraphics[width=\textwidth]{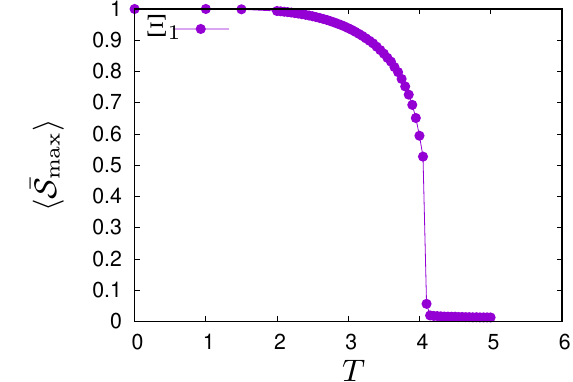}
	\end{subfigure}
	\begin{subfigure}[b]{0.329\textwidth} \caption{\label{F:345op_save_sigma_T_i} $K=5$}
	\includegraphics[width=\textwidth]{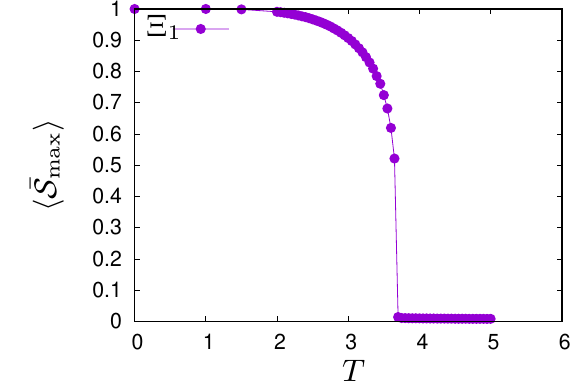}
	\end{subfigure}
	\begin{subfigure}[b]{0.329\textwidth} \caption{\label{F:345op_save_sigma_T_j} $K=3$}
	\includegraphics[width=\textwidth]{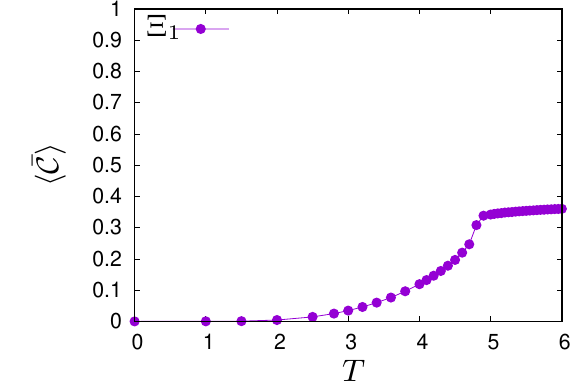}
	\end{subfigure}
	\begin{subfigure}[b]{0.329\textwidth} \caption{\label{F:345op_save_sigma_T_k} $K=4$}
	\includegraphics[width=\textwidth]{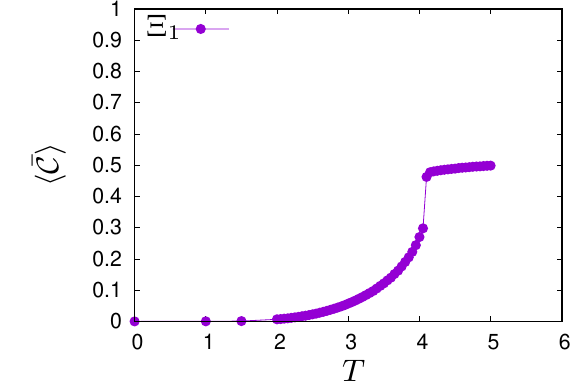}
	\end{subfigure}
	\begin{subfigure}[b]{0.329\textwidth} \caption{\label{F:345op_save_sigma_T_l} $K=5$}
	\includegraphics[width=\textwidth]{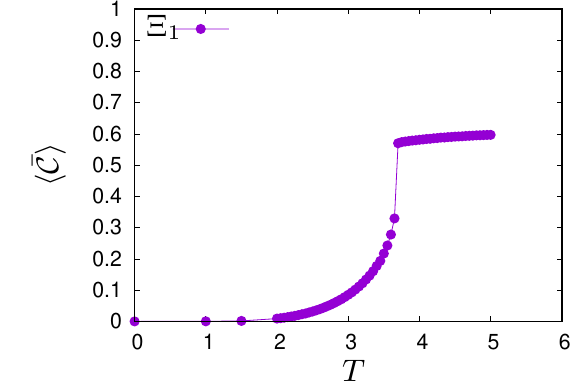}
	\end{subfigure}
\caption{\label{F:345op_save_sigma_T} The values of (a-c) $\langle\bar n_k\rangle$, (d-f) $\langle\sigma(\bar n_k)\rangle$, (g-i) $\langle\mathcal{\bar S}_{\text{max}}\rangle$ and (j-l) $\langle\mathcal{\bar C}\rangle$ for $L=40$ and $\forall i: p_i=s_i=0.5$ averaged over last $\tau=5000-100$ time steps.}
\end{figure*}

In Figs.~\ref{F:345op_save_sigma_T_a}-\ref{F:345op_save_sigma_T_c} and Figs.~\ref{F:345op_save_sigma_T_d}-\ref{F:345op_save_sigma_T_f} we plot $\langle\bar n_k\rangle$ and $\sigma(\bar n_k)$ for $K=3,4,5$, respectively.
As we can see in Figs.~\ref{F:345op_save_sigma_T_a}-\ref{F:345op_save_sigma_T_c} the majority $\langle\bar n_1(t=0)\rangle$ of holders of opinion $\Xi_1$ vanishes with increasing the social temperature $T$.
For critical social temperature $T\ge T_C$ all available opinions $\Xi_1,\cdots,\Xi_K$ in the system are equally occupied ($\langle\bar n_1\rangle=\langle\bar n_2\rangle=\langle\bar n_3\rangle\approx 33\%$ for $K=3$ and $\langle\bar n_1\rangle=\cdots=\langle\bar n_5\rangle\approx 20\%$ for $K=5$).

Again, vanishing of initially major opinion at $T=T_C$ coincidences nicely with maximal values of $\sigma(\bar n_k)$ as presented in Figs.~\ref{F:345op_save_sigma_T_d}-\ref{F:345op_save_sigma_T_f}. 
Similar critical behaviour may be observed in thermal evolution of the size of the larger cluster of actors sharing the same opinion $\langle\mathcal{\bar S}_{\text{max}}\rangle$ (see Figs.~\ref{F:345op_save_sigma_T_g}-\ref{F:345op_save_sigma_T_i}) and the total number $\langle\mathcal{\bar C}\rangle$ of clusters of actors sharing the same opinion (see Figs.~\ref{F:345op_save_sigma_T_j}-\ref{F:345op_save_sigma_T_l}). 
The increase of the number of clusters with increasing social temperature is also clearly visible in the Fig.~\ref{F:K3K6}.

\section{\label{S:disc}Discussion and conclusions}

In this paper we proposed multi-choice opinion dynamics model based on Latan\'e theory.
With computer simulation we show, that for multi-opinion version of the Nowak--Szamrej--Latan\'e model of opinion dynamics even without assigning numeric values for opinions we are able to observe phase transition similar to this occurring in two-state Ising-like models of opinion dynamics.

As we avoid signing a numerical values to possible opinions, we do not need to use the Likert-like scale~\cite{Likert1932} with possible Likert items as `1 = Strongly disagree', `2 = Disagree', `3 = Neither agree nor disagree', `4 = Agree' and `5 = Strongly agree'.
Likert scale falls within {\em the ordinal level} of measurement accordingly to the best known classification of scales of measurement by~\citet{Stevens677}.

Instead of signing a numerical values to possible opinions we deal with $K$ `colours' of opinions $\Xi_1,\Xi_2,\cdots,\Xi_K$ and probabilities of choosing these opinions given by Eqs.~\eqref{eq:E_ik}-\eqref{eq:P_ik}.
Please note that these `colours' are equally distanced to each other and none of them is better or worse than others.
Thus our scale of opinions corresponds to the {\em nominal level} of measurement~\cite{Stevens677}.
Please note, that term responsible for actors interactions with other actors who share the same opinions [Eq.~\eqref{eq:wplyw_ta_sama}] is not dissimilar to the Potts model~\cite{Potts_1952}, where phase transition is also observed.

As we do not assign numerical values $\xi_i$ to differentiate actors opinions we can observe the order/disorder phase transition in thermal dependence of $\langle\bar n_k\rangle$, $\sigma(\bar n_k)$, $\langle\mathcal{\bar S}_{\text{max}}\rangle$, $\langle\mathcal{\bar C}\rangle$.
The results of our simulations indicate that the critical temperature $T_C$ decreases with increasing the number of opinions $K$ available in the system. 
We conclude, that for opinion Nowak--Szamrej--Latan\'e model---with multi-choice of opinions and long-rage interactions among actors---the phase transition from ordered to disordered phase is also observed.

\begin{table}[!htbp]
\caption{\label{tab-TC} The values of critical social temperature $T_C$ for various number $K$ of opinion available in the system deduced from Figs.~\ref{F:2op_save_sigma_T} and~\ref{F:345op_save_sigma_T}.}
\begin{ruledtabular}
\begin{tabular}{lrrrr}
$K$   &   2 &   3 &   4 &   5 \\ \hline
$T_C$ & 6.1 & 4.7 & 4.1 & 3.6 \\
\end{tabular}
\end{ruledtabular}
\end{table}

\begin{acknowledgments}
This work was supported by the AGH-UST statutory tasks No. 11.11.220.01/2 within subsidy of the Ministry of Science and Higher Education.
\end{acknowledgments}

\appendix

\section{\label{app:manual} Manual}

Application was designed to show dynamic process of opinion formation. It allows to change parameters during simulation and tracking results.
Application consists of two main elements:
\begin{itemize}
\item control and results panels on the left,
\item simulation area on the right.
\end{itemize}
   
Control panels are divided into small windows. Each of them may be collapsed by clicking on the upper bar.
From the top there are the following panels:
\begin{itemize}
\item {\em Control}  allows to pause and restart simulation with applied parameters.
\item {\em Grid parameters}  contains following parameters: 
	\begin{description}
	\item[Height and Width] dimensions of the grid.
	\item[Number of opinion] $K$, number of available opinions.
	\item[Random opinion] indicates whether starting opinion of each actor is randomly chosen from $K$ allowed values or all actors have the same opinion.
	\item[Random parameters] when this option is selected then parameters $p_i$ and $s_i$ are randomly chosen from range $[0, 1]$ with uniform distribution. 
	\item[Parameter p, Parameter s] when \textbf{Random parameters} is not selected then all actors have the same fixed persuasiveness and supportiveness equal to this two values.
	\end{description}
Parameters located here can not be changed during simulation. After change user have to apply them by clicking button on bottom, after this new grid will be created and previous simulation will be lost.
\item {\em Simulation parameters} allows to change following parameters during the simulation:
	\begin{description}
	\item[External field] button which opens pop-up with $K$ numeric values. Changing these values introduce impact from outside. It allows to strengthen or weaken a particular opinion.
	\item[Temperature] social temperature $T$.
	\item[Distance function exponent] exponent $\alpha$ used in distance scaling function \eqref{eq:g}.
	\item[Delay] slider specifying pause time between steps of simulation. 
	\end{description}
\item {\em Leader parameters} allows to manually changing opinion and parameters $p_i$, $s_i$ of an actor in grid center. This allows to introduce a strong leader with persuasiveness and supportiveness higher than $1$.
\item {\em Results} numbers of actors with particular opinion.
\item {\em Chart} dynamically generated chart which shows the results over time. 
\end{itemize}
      
The simulation area is built of squares. Each square represents one actor. The color of square represents opinion. Height of the darker bars on the bottom of the square indicates values of parameters $p_i$ (on the left) and $s_i$ (on the right).  When simulation is paused user can click on actor to see impacts from each opinion and chance of changing opinion in next step.

\bibliography{this,opiniondynamics,ca,km}
\end{document}